\documentclass[aps,preprint,showkeys,showpacs,nofootinbib]{revtex4}%
\usepackage{amsfonts}
\usepackage{amsmath}
\usepackage{amssymb}
\usepackage{graphicx}%

\newcommand{\beq}{\begin{equation}}
\newcommand{\eeq}{\end{equation}}
\newcommand{\beqn}{\begin{eqnarray}}
\newcommand{\eeqn}{\end{eqnarray}}
\newcommand{\bearr}{\begin{array}}
\newcommand{\enarr}{\end{array}}

\newcommand{\eps}{\varepsilon}
\begin{document}


\title{An open-system approach for the characterization of  
spatio-temporal chaos\footnote{{\bf Running head:} Open system approach to spatio-temporal chaos}}
\author{P. Cipriani}
\author{A. Politi}
\thanks{Send proofs to A. Politi - Tel.:+39 055 23081 - Fax:+39 055 2337755}
\affiliation{Istituto Nazionale di Ottica Applicata, Largo E. Fermi 6, Firenze,
I-50125 Italy }

\date{\today}
\begin{abstract}
We investigate the structure of the invariant measure of space-time chaos by
adopting an ``open-system'' point of view. We consider large but finite windows
of formally infinite one-dimensional lattices and quantify the effect of the
interaction with the outer region by mapping the problem on the dynamical
characterization of localized perturbations. This latter task is performed by
suitably generalizing the concept of Lyapunov spectrum to cope with
perturbations that propagate outside the region under investigation. As a
result, we are able to introduce a ``volume''-propagation velocity, i.e. the
velocity with which ensembles of localized perturbations tend to fill volumes
in the neighbouring regions.
\end{abstract} 

\pacs{05.45.Jn, 05.45Df, 05.45Ra, 05.45Pq}
\keywords {High-dimensional Chaos; Fractals; Coupled map lattices; Numerical simulations of 
chaotic models.}

\maketitle

\section{Introduction}

Since the discovery of deterministic chaos, it has become clear that 
unpredictable behaviour can not only be the outcome of a stochastic dynamics, 
but also of a few nonlinearly-coupled degrees of freedom. Even more important
is to notice that these two classes of behaviour can be distinguished without
making any assumption on the underlying model. This is made
possible by embedding a supposedly recorded time series $u_j = u(t=j\tau)$
(where $\tau$ is the sampling time) into a space of dimension $L$ (i.e.,
introducing the vector $U_j^{(L)} = \{u_j,u_{j+1},\ldots u_{j+L-1}\}$) and
thereby estimating the fractal dimension $D(L)$. In stochastic processes,
$D(L)=L$, since the variables at different times are mutually independent (at
least below a certain threshold, which clearly depends on $\tau$). 
In low-dimensional chaos, when $L$ is
increased, $D(L)$ saturates to a finite value because of the functional
dependence among the variables. This is the starting point of nonlinear
time-series analysis, an ensemble of tools, developed in the last years to
reconstruct a deterministic model starting from raw data,\cite{KS}.

More subtle is the difference between stochastic and deterministic signals, 
when the latter ones arise from a high-dimensional dynamics (as in space-time 
chaos), since $D(L)$ increases with $L$ in both cases. In this context, it is
still possible to distinguish between the two classes of behaviour, provided
that a more refined analysis, based on the effective, or coarse-grained, 
dimension $D_c(\eps,L)$  is developed, where $\eps$ represents the resolution of the
coarse graining.

In fact, in stochastic systems, all variables turn out to be mutually
independent below a fixed threshold $\eps_s$ that is independent of the
embedding dimension $L$, so that $D_c(\eps,L)$ is essentially equal to $L$ for
$\eps < \eps_s$. On the other hand, in space-time
chaos, it has been conjectured that new degrees of freedom appear only at the
expense of progressively decreasing the observational threshold. A loose
explanation for this behaviour is based on the observation that the degrees
of freedom associated to far-away regions are almost decoupled from the
evolution in the observation point \cite{P85}. Although this statement looks very
reasonable, it is not at all easy to substantiate it with solid arguments.
Indeed, very little progress has been made in the last decade and what is
known is still mostly based on heuristic arguments.
 
In this paper, we attack the problem of characterizing space-time chaos by
constructing a spatial rather than a temporal embedding, i.e., by referring to
the hypothetical signal $u_j = u^j(t)$, where the time $t$ is large enough to
ensure convergence to the attractor, while $j$ now labels the lattice sites
(for the sake of simplicity we limit ourselves to considering one-dimensional
lattices). 
Accordingly, the effective dimension $D_c(\eps,L)$ now counts the
number of degrees of freedom that can be resolved, with resolution $\eps$,
in a window of length $L$ 
embedded in a supposedly infinite system. This problem is conceptually
equivalent to the previous one, the main difference being that time and
space axes have been exchanged. It is precisely this difference that allows
us using the more standard tools developed to investigate the invariant measure
of chaotic systems.

In the investigation of space-time chaos, closed systems of finite length
$L$ are typically considered. In such a context, the classical tools for the
characterization of chaotic dynamics  such as the Lyapunov exponents
describing the evolution of infinitesimal perturbations, can be effectively
implemented. As a result, it has since long been recognized that a limit
Lyapunov spectrum does exist for $L\to \infty$, thereby inferring (from
Kaplan-Yorke and Pesin formulae) the extensivity of space-time chaos
\cite{G89}: in fact, both the fractal dimension and Kolmogorov-Sinai entropy
are proportional to the system volume (length, in one dimension).

At variance with this approach, here we adopt an ``open''-system point of view,
i.e. the finite window of length $L$ is part of an infinite system whose
evolution is taken into account as well. In a sense, the two methods are
reminiscent of the microcanonical and canonical ensembles of statistical
mechanics; the open-system approach is indeed a possible way (though not the
most effective one) to perform canonical simulations of Hamiltonian systems.
Unfortunately, at variance with equilibrium statistical mechanics, here there
is not a prescription such as the Boltzmann weight to estimate, a priori, the
probability of each configuration. All we have at our disposal is the nonlinear
dynamical law with the additional difficulty (in comparison to the
``closed''-system approach) of having to deal with perturbations propagating
from the outer to the inner region and viceversa. 
 
Collet and Eckmann \cite{CE99} have developed a method that more than any other 
has inspired us in developing the approach outlined in this paper. Their idea
is that the structure of the invariant measure in a window of fixed length $L$
at a given observational scale can be described by following in time the
convergence of a suitable ensemble of initial conditions towards the attractor.
In fact, such an evolving ``cloud'' of points provides a natural covering of
the attractor that becomes increasingly sharp upon time evolution. Turning the
time dependence of infinitesimal ellipsoids into a dependence on the
observational resolution $\eps$ is a possible way to explain the
Kaplan-Yorke formula in standard finite dimensional systems. However, the 
extension of such ideas to open systems requires that the very concept of
Lyapunov spectrum be revisited so as to describe perturbations that evolve
also outside the region that is currently monitored. We shall see that 
a meaningful characterization of the perturbation dynamics can be obtained
only by realizing that the problem involves two scaling parameters that must
be simultaneously let diverge to infinity: the window length $L$ and the
evolution time $T$. In fact, we end up introducing a Lyapunov spectrum that,
besides depending, as usual, on the ratio of the label of each exponent by the
length $L$, depends also on $T/L$. This is not simply an additional technical
difficulty, but the key element that allows mapping the characterization of
pertubation evolution onto the characterization of the invariant measure in
spatially extended systems. In fact, expressing the variable $T$ in terms of the
observational resolution allows sheding further light on the dependence of the
effective dimension $D_c(\eps,L)$ on both $L$ and $\eps$. While our analysis does
confirm the functional dependence conjectured in \cite{T93,K94}, we find
different expressions for the coefficients.

A brief summary of the known results is presented in Sec.~II. Sec.~III is
devoted to a detailed justification of the Kaplan-Yorke formula by following an
approach that can be most easily extended to open systems. A relevant result of
our analysis is the above mentioned extension of the concept of Lyapunov
spectrum: this issue is the core of Sec.~IV, where we also illustrate the
implementation of the method in various classes of coupled map lattices.
Finally, in Sec.~V, we  discuss the implications of the Lyapunov analysis on
the fractal properties of the invariant measures in open systems and briefly
discuss the further corrections expected to arise from the boundaries.

\section{The state of the art}

As already anticipated in the introduction, we aim at characterizing the
structure of the invariant measure of space-time chaos.
For the sake of simplicity, most of the analysis will be restricted to 
(one dimensional) lattice
systems. A detailed description of the scaling properties of a given set 
is contained in the ``effective'' dimension
\begin{equation}
{\cal D}_c(\eps) = - \frac{d H}{d \ln \eps} \quad ,
\label{deff}
\end{equation}
where $H(\eps)$ is the entropy of the set covered with boxes of size $\eps$, 
$H = -\sum p_i \ln p_i$, $p_i$ being the probability of 
each box\footnote{Rigorously speaking, one should refer to the optimal covering.
We implicitly assume to have made such a choice}. 
As shown by Renyi long ago, different definitions of entropy can
be given, by replacing the logarithmic average of $p_i$ in the $H$ expression
with averages of its moments. As, in general, there are tiny differences among
the various entropies, we will always refer to $H$ without specifying which
average is being taken.
 
To our knowledge, the concept of a resolution-dependent entropy was first
introduced by Kolmogorov and Tikhomirov \cite{KT}, who called it $\eps$-entropy. 
In principle, an $\eps$-dependent dimension is an ill-defined concept, since
it is not invariant upon change of space parametrization. For this reason, one
has to take the limit $\eps \to 0$. In fact, only in this limit, ${\cal D}_c(0)$ 
becomes a dynamical invariant that is strictly related to other dynamical 
invariants such as the Lyapunov exponents. However, besides the asympotic value
${\cal D}_c(0)$, also the possibly slow dependence on $\eps$ at high resolutions
can be universal, thus conveying meaningful information about the structure
of the set of interest. This is precisely one of the reasons why 
$\eps$-entropies have been introduced to quantify the cardinality of different
classes of functions (such as, e.g., the entire functions) \cite{KT,CE99}.

In the study of spatially extended systems, one is faced with the difficulty 
of accounting for the dependence on a further parameter besides $\eps$, namely
the system size $L$. However, previous studies of closed systems have clearly revealed
that, for sufficiently large $L$, the coarse grained dimension is still
an extensive quantity: \cite{G89}
\begin{equation}
 D_c(0,L) \propto dL
 \label{first}
\end{equation}
where $d$ can be interpreted as the dimension density (i.e. the contribution
to the dimension per lattice site) that can be determined from the Kaplan-Yorke
formula (see next section).

In open systems, it is instead clear that $D_c(0,L) = L$, since the
infinitely many degrees of freedom ruling the outer part of the chain act as
a sort of stochastic source \cite{TPPA91,PP92}. In order to understand how the two
results can be reconciled, it is necessary to investigate, in the case of open systems,
the simultaneous dependence on both $\eps$ and $L$ (within the closed system approach, the
resolution does not play an important role -- see next section). As long
as $\eps$ and $L$ are respectively small and large enough, the scaling
dependence on the two parameters is expected to be universal. 

The first consistent conjecture about this problem was formulated by
Korzinov and Rabinovich \cite{K94}, 
\begin{equation}
  D_c (\eps,L) = dL - \frac{vd^2}{\eta} \ln \varepsilon - A\ ,
\label{korz}
\end{equation}
where $d$ is again the dimension density,
$\eta$ is the Kolmogorov-Sinai entropy density \cite{G89}, $v$ is the propagation
velocity of disturbances, and $A$ is a non-better-specified parameter.
This equation parallels the analogous expression proposed by Tsirimg for the symmetric
problem of the dimension of a scalar time series recorded at a single 
point \cite{T93}.  
One can notice that Eq.~(\ref{korz}) allows reconciling the apparently
contradictory expectations for closed and open systems. In fact, if the limit
$\eps \to 0$ is taken before the limit $L \to \infty$, $D_c/L$ diverges (though,
in reality, it could not become larger than 1 - this inconsistency is due to the
perturbative character of the above formula); if the order of the limits is reversed,
then, $D_c/L$ converges to the expected (closed system) value $d$.

Unfortunately, the derivation of the above formula depends on several 
assumptions that cannot be directly checked. Moreover, it is rather unlikely 
that more accurate numerical simulations will provide clean enough data to 
draw definite conclusions. It is therefore compelling to make some progress on
the theoretical side, even at the expense of introducing strong simplifications.
This is the route already undertaken in Ref.~\cite{OHK98}, where the limit case of
weakly  coupled maps has been considered. 

If the invariant measure is assumed to cover a linear subspace, it is possible
to obtain a detailed description of it \cite{PW99}. In fact, under this
approximation, it is possible to implement global methods such as
singular-value decomposition (SVD) technique.
In general, the usefulness of SVD is limited by the presence of nonlinearities 
that induce bendings which, in turn, do not permit extracting information
about the local thickness of a given set. Let us, for instance, imagine a
slightly bended segment embedded in a two-dimensional plane. SVD will 
tell us that the set of points belonging to the segment is characterized by
two non-zero orthogonal widths even though the set is strictly one-dimensional. 
If one can restrict the discussion to linear subspaces, this problem does not
arise and a global method like SVD can be effectively used to extract local 
information. In Ref.~\cite{PW99}, it has been assumed that the invariant measure
of the infinitely extended system is the linear superposition of a subset of
all possible modes in a given basis (e.g., all Fourier modes with wavenumber
smaller than a prescribed threshold). The fraction $d$ of the active modes is
the dimension density of this given space of functions. The corresponding
problem of characterizing the projection onto a space of finite width $L$ can
be addressed by computing the eigenvalues of a suitable correlation matrix. 
As a result, it has been found that the observational resolution $\eps$ and
the effective dimension $D_c$ are connected by the scaling relation
\begin{equation}
  \ln {\eps} = - L F(D_c/L) ,
\label{scaling}
\end{equation}
where the function $F(x)$ is identically zero for $x<d$, while it increases 
monotonously for $x>d$, starting with a finite slope. Solving the above equation
with respect to $x=D_c/L$ and expanding the resulting inverse function, 
$F^{-1}(\ln{\eps})$ for small values of $\ln{\eps}$, one finds the perturbative expression 
\begin{equation}
  D_c (\eps,L) = dL - \frac{\ln \eps}{\beta_1} -
    \frac{\beta_2}{\beta_1^3L}(\ln \eps)^2  ,
\label{deff2}
\end{equation}
that has the same structure of Eq.~(\ref{korz}). Thus, Eq.~(\ref{scaling}) 
appears as its natural extension to arbitrarily small scales.

The main drawback of this approach is the absence of any dynamics. In
particular there is no way to link the function $F(x)$ to dynamical invariants
such as Lyapunov exponents. 

A more suitable starting point is represented by the study of the complex
Ginzburg-Landau equation performed by Collet and Eckmann \cite{CE99}. 
They have rigorously proved that 
\begin{equation}
D_{CE}(\eps,L) = B_0L + B_1/\varepsilon^2
\label{GLCE}
\end{equation}
represents an upper bound to $D_c$. On the one hand, this formula confirms
the existence of a leading term that is proportional to the system size.
On the other hand, the above expression differs from the previous ones for 
what concerns the leading correction, that diverges faster than logarithmically
for $\eps \to 0$. 
The question whether such a difference is to be attributed to the continuity 
of the space variable (and thus to the possibly larger number of degrees of 
freedom) or it is due to technical difficulties in improving the upper bound 
cannot be easily answered. On the basis of the results here presented
we argue that, for lattice systems, the upper bound Eq.~(\ref{GLCE})
can be improved, while for spatially continuous flows, the situation is yet
unsettled.

The idea behind the derivation of Eq.~(\ref{GLCE}) is the same that allows 
proving the Kaplan-Yorke formula for standard finite-dimensional attractors:
given a set of boxes that cover the attractor, we can obtain finer
coverings by simply letting each box evolve in time. We summarize the idea
in the next section, as it will be useful for the generalization to open
systems.

\section{The Kaplan-Yorke formula}

In this section, we discuss a method that allows extending the Kaplan-Yorke
formula to open systems. It is both useful and necessary to start from the
simple context of a 2d chaotic map (such as, e.g., the H\' enon map). Let us cover
the attractor with a square $S_0$ of size ${\mathcal O}(1)$ (in general it will
be a hypercube) and let us denote with $S_t$ its image after $t$ time steps.
$S_t$ provides a covering of the attractor at all times, even though stretching
and folding transform it into a long and thin sausage. It is natural to divide
$S_t$ into boxes of size equal to its average width (for the sake of simplicity,
we do not take into account multifractal fluctuations, that would not anyhow
modify the following scaling arguments)
\begin{equation}
\eps = \delta_2(t) = \exp\{\lambda_2t\}
\end{equation}
where $\lambda_2$ is the second, negative, Lyapunov exponent. This is the crux
of our argument, as it suggests how time evolution spontaneously
introduces an increasing resolution in phase-space. In fact, we can now
estimate the fractal dimension ${\cal D}(\eps)$ from the number of boxes $N(\eps)$ 
needed to cover $S_t$ with resolution $\eps$,
\begin{equation}
{\cal D}(\eps) = -\frac{\ln N}{\ln \eps} =  -\frac{\ln({\rm e}^{\lambda_1 t}/\eps)}{\ln \eps} =
  1-\frac{\lambda_1}{\lambda_2}
\label{ky1}
\end{equation}
where $\lambda_1$ is the positive Lyapunov exponent (in the philosophy of
neglecting multifractal corrections, we do not distinguish between the positive
Lyapunov exponent and the topological entropy). $S_t$ provides a meaningful
covering of the attractor only if the dynamics is invertible, otherwise
the above equation would represent only a (possibly rough) upper bound.
Eq.~(\ref{ky1}) is nothing but the well known Kaplan-Yorke formula in 2d maps.
It is worth recalling that although this derivation is rather sketchy, the
equality is rigorous, provided that ${\cal D}$ is interpreted as the information
dimension \cite{ER85}. 

This approach can be extended to higher dimensional maps, but it requires
an additional assumption that seems to be generally valid, although not
universally correct. Before discussing the most general case, let us first add
a decoupled, contracting direction to the previous system, as this case helps
clarifying the difficulties that arise in higher dimensions. The length of $S_t$
remains unchanged, while its transversal section becomes an ellipse with two
semi-axes of length $\delta_2 = \exp(\lambda_2 t)$ and 
$\delta_3 = \exp(\lambda_3 t)$, respectively ($\lambda_3$ being the additional,
negative, Lyapunov exponent). The question now consists in choosing the size
$\eps$ that allows an optimal covering of $S_t$. In this special case, we
know a priori that $N(\eps)$ must not change, since we have not modified the
attractor itself. If we choose $\eps = \delta_3$, the resulting $N(\eps)$ can be
either much larger or smaller than before, depending on the relative size of
$\delta_2$ and $\delta_3$. The error of this choice is that the hidden structures
not yet resolved at time $t$ are contained only along the second direction and
thus, we must fix $\eps = \delta_2$. 

In this case, the right result has been obtained because, in a sense, we
already knew the solution. Let us now refer to a generic finite-dimensional
system and let $N_p$ denote the number of positive Lyapunov exponents. An
initial hypercube $S_0$ with edge-length ${\mathcal O}(1)$ covering the
attractor is stretched along the unstable directions and contracted along the
stable ones. Since the attractor is bounded along all
directions, the ``excess'' of length that is continuously produced along the
unstable directions must be folded along the contracting ones. The crucial
point is the assumption that folding generically proceeds from the
least to the most contracting directions. Technically, this is equivalent to
assuming that all contracting directions are filled until $\Lambda_k \doteq
\sum_{l=1}^k \lambda_l<0$ (where the Lyapunov exponents are implicitly ordered
from the largest to the most negative one), while the remaining most stable
directions do not contribute (like the third direction in the above example).
Accordingly, the ``right'' box-size to be adopted in the partitioning process
of $S_t$ is
\begin{equation}
\eps = \exp(\lambda_{m}t)
\label{timres}
\end{equation}
where $m$ is the minimum $k$-value such that $\Lambda_k <0$. From the
 corresponding
number of boxes of size $\eps$ needed to cover $S_t$, it is readily found that 
\begin{equation}
{\cal D}_{KY} =  m-1 + \frac{\Lambda_{m-1}}{\lambda_m} .
\label{KYgeneral}
\end{equation}
This is the general form of the Kaplan-Yorke formula. 

In spatially extended
systems of large length $L$, the Lyapunov exponent depends on the index $l$ and
$L$ only through the scaling variable $\rho = l/L$, i.e. 
$\lambda_l = \lambda(\rho)$.
Accordingly, by neglecting the fractional correction in Eq.~(\ref{KYgeneral}),
the dimension can be written as 
\begin{equation}
{\cal D}_{KY} = d_{KY}L
\label{KYdens}
\end{equation}
where $d_{KY}$, implicitly defined by the constraint 
$\int_0^{d_{KY}} d\rho \lambda(\rho)=0$, can be interpreted as a density of
dimension (see also Eq.~(\ref{first})) \cite{G89}.

All of the above discussion can be summarized stating that the number of boxes
needed to cover an attractor with resolution $\eps$ can be determined by letting
a ball of size ${\mathcal O}(1)$ evolve until its width along the $m$-th direction
(determined by the vanishing of $\Lambda_m$) is equal to $\eps$ itself. 
In other words, Eq.~(\ref{timres}) is the core of the argument, as it allows 
transforming the dependence on $t$ into a dependence on the resolution.

In the above discussion we have implicitly assumed that each Lyapunov exponent
is equal to its asymptotic value, independently of $t$ (and thus of the
resolution). As long as each Lyapunov exponent exhibit a slow dependence on
time, all of the above discussion still applies, with the difference that the
r.h.s. of Eq.~(\ref{KYgeneral}) depends on $\eps$ (through the hidden
dependence of the $\lambda$'s on $t$). In other words we see that the
Kaplan-Yorke formula is ready to account not only for the asymptotic value of
the fractal dimension but also for possible dependencies on the observational
resolution. This is precisely what happens in the case of open systems.\\

Before proceeding further in this direction, we need to introduce some
notations: let ${\bf x}_\parallel$ and ${\bf x}_\perp$ denote two vectors
defining the state variable on each lattice site, within, respectively,
outside, the window of interest $W_L$. As in the previous discussion, the aim
is to infer the fractal properties of the attractor from the density 
$P(t,{\bf x}_\parallel,{\bf x}_\perp)$ at time $t$  (the initial condition
$P(0,{\bf x}_\parallel,{\bf x}_\perp)$ being a constant distribution in a
hypercube of radius order 1 that contains the attractor). The probability
density can be usefully rewritten as
\begin{equation}
P(t,{\bf x}_\parallel,{\bf x}_\perp) = \int d {\bf x}_\perp^0 
    Q(t,{\bf x}_\parallel,{\bf x}_\perp|{\bf x}_\perp^0) 
    P_\perp({\bf x}_\perp^0)
\label{int}
\end{equation}
where $Q(t,{\bf x}_\parallel,{\bf x}_\perp|{\bf x}_\perp^0)$ denotes the
probability density at time $t$ conditioned to the initial state
${\bf x}_\perp^0$ of the external variables, while $P_\perp({\bf x}_\perp^0)$
represents their distribution. The integral over the ``hidden'' variables
${\bf x}_\perp^0$ represents the first relevant difference with the previous
case: the ignorance about their values contributes to dressing the probability
density. We will discuss a bit this problem in the last section.

Anyhow, even disregarding the effect of the integral in the above equation,
the problem we have to deal with is more complex than the previous one.
In fact, even if we consider
initial conditions $({\bf x}_\parallel^1,{\bf x}_\perp^0)$, 
$({\bf x}_\parallel^2,{\bf x}_\perp^0)$ that differ only inside $W_L$, the
mutual difference does not remain confined to $W_L$, but rather spreads and
propagate in the outer regions. Accordingly, we are faced with the problem of
defining, in this context, the Lyapunov spectrum in a meaningful way. This is
the goal of the next section.

However, before discussing the generalization of Lyapunov spectra to open
systems, we briefly present a heuristic explanation of Eq.~(\ref{GLCE}),
since its derivation is based on the idea that a characterization of the
invariant measure over increasingly fine scales can be obtained from the
evolution of a set $S_0$ covering the attractor. More precisely,
in Ref.~\cite{CE99}$, W_L$ has been split into two parts: the bulk $B$, where the
effect of the external degrees of freedom can be neglected (over the time $t$),
and the boundary $b$, where propagation must be, instead, taken into account.
An upper bound to the number of boxes needed to cover the attractor has then
been estimated as the product of the number of boxes needed to cover $B$, times
the number of boxes needed to cover $b$. In the computation of both quantities
one is faced with the difficulty of dealing with a continuous spatial variable.
Such a problem has been solved by introducing a proper discretization $\delta$.
As a result, it turns out that in the bulk, $\delta = {\mathcal O}(1)$,
while, inside $b$, it is $\delta = {\mathcal O}(\eps)$. The reason for the difference is
that in the bulk, nearby configurations change their mutual distances only as a
result of local instabilities that are of order 1. On the contrary, in $b$,
the difference may also grow due to the propagation of uncontrolled
perturbations from the boundaries. Accordingly, the fractal dimension is
basically proportional to the number of lattice points introduced in the
discretization processs: in the first and second term of the r.h.s.
of Eq.~(\ref{GLCE}), one can recognize the contributions arising from the bulk
and the boundaries, respectively. The latter one has size ${\mathcal O}(1/\eps^2)$, 
because the length of the boundary is estimated to be ${\mathcal O}(1/\eps)$. 
Since in lattice systems there is a natural spacing, an extension of this reasoning
to that context would lead to a correction term of order ${\mathcal O}(1/\eps)$, to
be confronted with the logarithmic correction predicted by
Eqs.~(\ref{korz},\ref{deff2}). In the last section we will briefly discuss the
possible reasons of such a discrepancy.

\section{Lyapunov spectra of open systems}

It is several years that the concept of convective Lyapunov exponent has been 
successfully introduced to describe how perturbations spread and grow. This 
is done by measuring at time $t$ the amplitude $\delta x^i(t)$ of an
initially $\delta$-like perturbation ($\delta x^i(0) = \delta^i_0$) and
determining its growth rate in a frame moving with velocity $v$ \cite{DK87},
\begin{equation}
 \Lambda_c(v) = \lim_{t\to \infty} \frac{\ln |\delta x^i(t)|}{t}\ ,
\end{equation}
where $i = vt$.

It is known that whenever the spatial left-right symmetry is
not broken, the maximum value for $\Lambda_c(v)$ is obtained for $v=0$ and it coincides
with the standard maximum Lyapunov exponent. Upon increasing the velocity, the
convective exponent decreases and becomes negative for $v > v_c$, to indicate
that only perturbations moving with a velocity slower than a critical velocity
can be sustained.

One might imagine to generalize this procedure, by looking not just
at the amplitude of a single perturbation but to the volumes spanned by a
finite number of perturbations, very much in analogy to what done for computing
Lyapunov spectra in closed systems. However, if we entirely follow the standard
approach, we are bound to conclude that all the convective exponents that can
be associated to a given velocity coincide with the maximal value. The reason
for this conclusion is that, on the one hand, the finite number of
perturbations that are followed in time visit a space of increasing
(eventually infinite) dimension. On the other hand, the existence
of a limit Lyapunov spectrum means, as discussed above, that $\lambda_l$ is a function
of $\rho = l/L$ alone, where $l$ denotes the $l$-th exponent, $L$ being the
length of the system. Since in the above setup, the number of exponents is
fixed and equal to $L$, while the system size increases with time, one is
basically computing an increasingly thin portion of the Lyapunov spectrum and,
eventually, all Lyapunov exponents become equal to the maximum one. Stated otherwise,
by implementing the usual closed systems approach, without modifications, we
compute the invariant Lyapunov spectrum, $\lambda(\rho)$, restricted to a range
$0<\rho\le \rho_{max}\doteq L/L_{eff}$, where $L_{eff}$ is the effective dimensionality of
the space explored, which increases indefinitely with time, thus implying that
$\rho_{max}\to 0$ when $t\to\infty$.

Although this is an inescapable conclusion, we now show that meaningful
results can be obtained even if perturbations are followed for a finite time.
A priori, one might think that computing over a finite time implies that the
corresponding quantity is ill-defined, because it would depend on the choice of
coordinates. However, in so far as time is finite but arbitrarily large, this
objection does not apply. This is for instance the case of the so-called
multifractal analysis of low-dimensional chaos. 
In the present context there are two scaling parameters to deal with: the
length $L$ and the time $T$. We eventually want to let both diverge to infinity.
The standard approach adopted in the literature consists in first letting
$T$ diverge to infinity (this allows determining the Lyapunov spectrum of a
finite system) and then taking the thermodynamic limit $L \to \infty$.

The problem of choosing the most appropriate order in the problem at hand is
very similar to the problem mentioned in the introduction about the order of
the two limits $L \to \infty$ and $\varepsilon \to 0$ for a meaningful
definition of fractal dimension in open systems. We propose here to let $T$
and $L$ diverge simultaneously, with fixed ratio,
\begin{equation}
g =  T/L\ ;
\end{equation}
{\it i.e.,} given a subsystem of length $L$ (embedded in a formally infinite 
chain), we let perturbations evolve for a time $T = gL$. Our claim is that the
corresponding spectra converge, in the limit $L \to \infty$, to a specific
shape that depends only on $g$. 

In order to be more precise, we start defining
all the quantities of interest. Consider $L$ independent vectors
$\{{\bf u}_n(t)\},\ (n=1,\ldots,L)$. Each ${\bf u}_n$ denotes a perturbation
initially restricted to a subchain of length $L$ of our (virtually) infinite
chain: (that is, $\forall\, n=1,\ldots,L$, it is 
$u^{(i)}_n(0)=0$ for $i\le 0$ and $i>L$, where $u^{(i)}_n$ stands for the $i$-th
component of the $n$- perturbation vector). 
Let us introduce the projection operator, ${\bf P_L}$
\begin{equation}
\{{\bf P_L}{\bf u}_n(t)\}^{(i)} = \begin{cases} u^{(i)}_n(t)  & {\rm if}
                                         \quad 0 < i\le L \cr
                                       0 & {\rm otherwise}\ . \end{cases}
\end{equation}
Let then evolve the $L$ vectors $\{{\bf u}_n(t)\}$  up to a time
$T=gL$, and determine the volumes spanned by the projection of $k$ such vectors
(with $k=1,\ldots,L$),
as one normally does in the computation of standard Lyapunov exponents. 
In this way we can compute the Lyapunov spectrum over a time $T$ in a spatial 
window of size $L$.

This approach applies to any one-dimensional system, irrespective of the
continuity/discreteness of the space and time variables. However, for the sake
of computational simplicity, we shall restrict ourselves to consider coupled
map lattices. In particular, we will mainly refer to the typical coupled map
lattice with diffusive coupling,
\begin{equation}
x^i(t+1) = f\left[ (1-2\eps)\,x^i(t) + \eps [x^{i-1}(t) + x^{i+1}(t)]\right] \ ;
\end{equation}
where $\eps$ is the coupling constant and $f(y)$ is a map of the unit interval
onto itself. 

In particular, we start considering Bernoulli maps, where
\beq
f(y) = a (y -[y]) 
\eeq
and the square brackets denote here the integer part.

\begin{figure}[ht]
\includegraphics*[width=13cm]{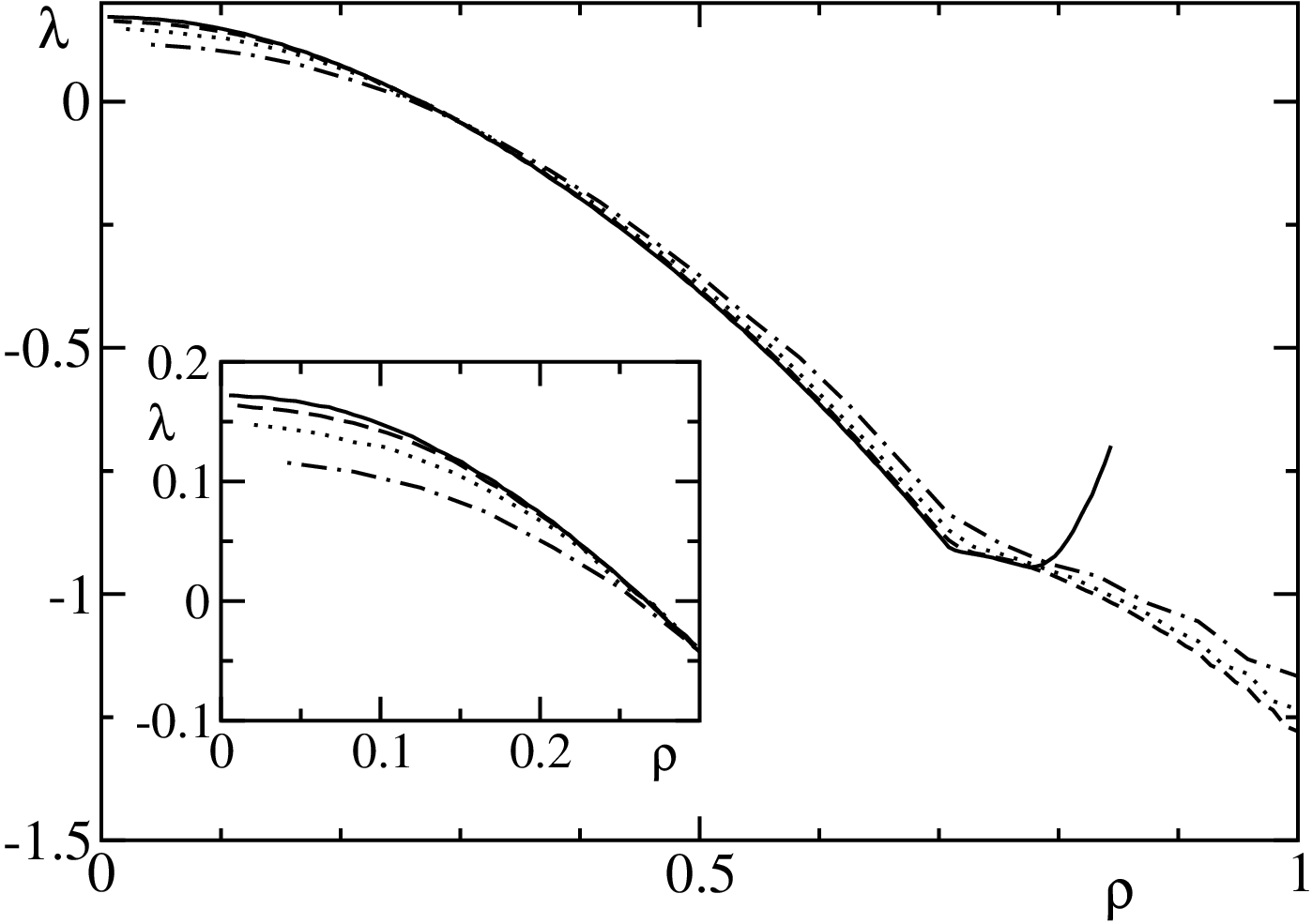}
\caption{Open-system Lyapunov spectra of Bernoulli maps with $a=1.2$ and $g=1$. 
Dot-dashed, dotted, dashed and solid curves correspond to $L=24$, 48, 96 and
192, respectively. The last part of the solid curve is truncated because it
is numerically unreliable. In the inset, an enlargement of the region around 
the maximum is reported.}
\label{conv}
\end{figure}

In Fig.~\ref{conv}, the spectra corresponding to the same ratio $g=T/L=1$, but 
different lengths, have been plotted. One can clearly see a convergence towards
an asymptotic limit. The large deviation observed in the bottom part of the
solid curve is due to numerical inaccuracies. Indeed, these spectra, that are
obtained by letting the perturbations evolve in the whole space and then
projecting them onto the window of interest, require an increasing accuracy
for increasing elapsed time. For $L=192$ even FORTRAN extended accuracy
(equivalent to approximately 30 digits) is no longer sufficient. Anyway, the
inset clearly confirms the tendency to converge towards a well defined
asymptotic shape, so that one can meaningfully introduce the concept of
open-system Lyapunov spectrum $\lambda(\rho,g)$ (OSLS).

\begin{figure}[tpb]
\includegraphics*[width=13cm]{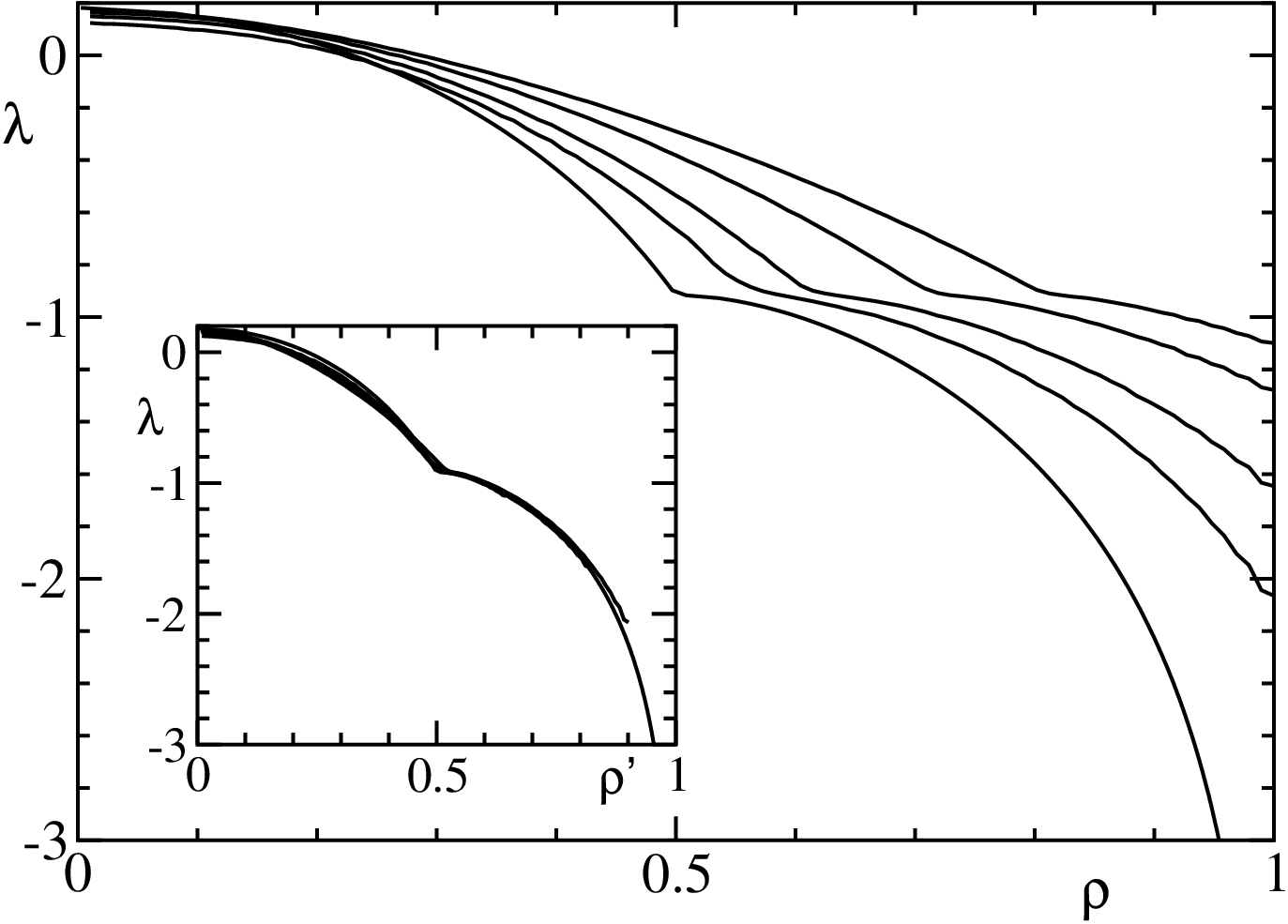}
\caption{Open-system Lyapunov spectra of Bernoulli maps with $a=1.2$ and
$L=96$. From bottom to top, the curves correspond to $g=0$, 1/4, 1/2, 1, and 3/2,
respectively. In the inset, the same spectra are plotted after
rescaling the $\rho$ axis according to the argument discussed in the text.}
\label{allbern}
\end{figure}

The spectra reported in Fig.~\ref{allbern} correspond instead to different values
of $g$ for a fixed length $L$. 
In the limit $g\rightarrow 0$, the effects of propagation outside the
initial window are negligible, so that $\lambda(\rho,g=0)$ reduces to the
standard Lyapunov spectrum. In fact, the lowermost curve corresponds to the
analytically known expression for the standard Lyapunov spectrum. 
Upon increasing $g$, the OSLS increases and for $g \to \infty$, we expect it
to flatten around the maximum Lyapunov exponent. In the inset of
Fig.~\ref{allbern}, we have suitably rescaled the $\rho$ axis. The rather good
data collapse suggests that the propagation of perturbations does not modify
the spectrum structure, but simply leads to an expansion of the $\rho$ scale. 
Indeed, from the
definition of $\rho=l/L$, one can notice that the scale is controlled by the 
window length $L$.

However, as anticipated above, in the case of open-system simulations, 
the space covered by each perturbation increases with time, so that 
it is reasonable that the label $l$ should be more properly scaled to some 
effective length $L_{eff}$ rather than to the initial length $L$. 
Furthermore, it can be conjectured that the
effective length increases linearly in time as $L_{eff} = L + 2vT$, with some
velocity $v$ (the factor 2 is included to account for the growth on both sides
of the window).  Therefore, the ``right" scaled variable should be
$\rho' = l/[L+2vT]= \rho/(1+2gv)$, and, asymptotically, one expects that
$\lambda(\rho,g)$ converges to some $\tilde{\lambda}(\rho')$. 
The good data collapse observed in 
the inset Fig.~\ref{allbern} points in this direction, although, from the raw
data we cannot exclude that the apparent hyperscaling is only an approximation.
Indeed, the finite size corrections affecting the various spectra are of the
same order as the deviations among the various curves. 

\begin{figure}[tpb]
\includegraphics*[width=13cm]{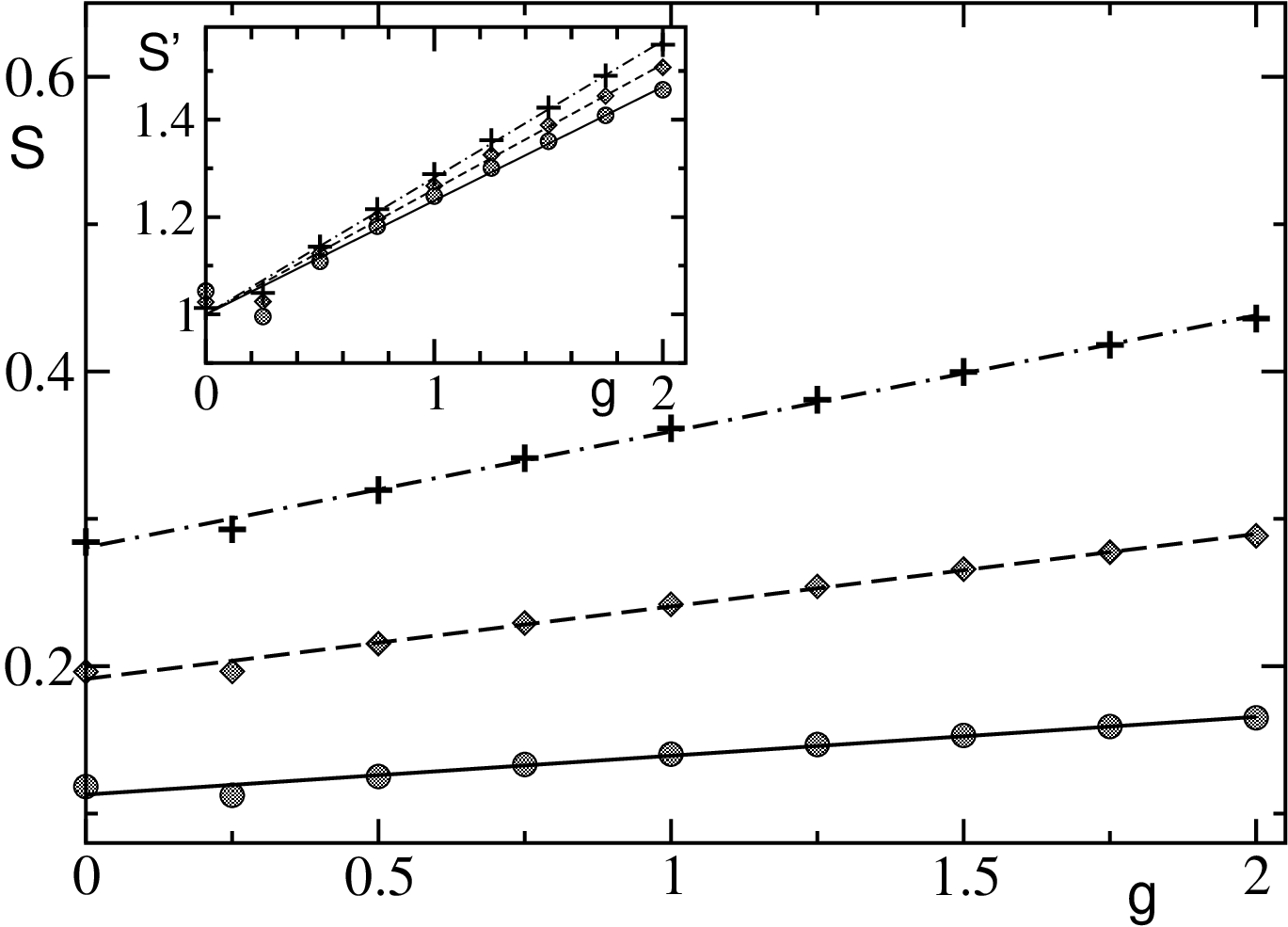}
\caption{The pseudo entropy $S(\lambda_s,g)$  defined in Eq.~(\ref{snodeep}) versus $g$ for
$\lambda_s = -0.7$, $-0.5$ and $-0.3$ (from top to bottom), for 
Bernoulli maps with $a=1.2$ and $L=96$. In the inset
the same data are plotted, with the same symbols, after rescaling, giving
the quantity $S\,'(g)$ defined in the text.}
\label{hKS}
\end{figure}

In order to study the dependence of the OSLS on $g$ in a more quantitative way,
we proceed as follows. Given a spectrum $\lambda(\rho,g)$, we fix a threshold
$\lambda_s$ and compute the quantity:
\begin{equation}
S(\lambda_s,g) \doteq \int (\lambda(\rho,g)- \lambda_s)\, d\rho  ,
\label{snodeep}
\end{equation}
where the integral is restricted to the interval of $\rho$-values where
the integrand is positive. For $\lambda_s=0$, $S$ reduces to the well known
Kolmogorov-Sinai entropy. The dependence of $S$ on $g$ can be observed in
Fig.~\ref{hKS} for three different thresholds. The essentially linear behaviour
confirms that the increase is due to a propagation process, since the effective
length increases linearly with time, {\it i.e.,} with $g$. If only one propagation velocity
$v_s$ is present in the evolution, each curve should increase as
\begin{equation}
 S(\lambda_s,g) = S(\lambda_s,0)(1+2v_s g) \quad .
\end{equation}
Therefore, the curves
obtained for different values of $\lambda_s$ should,
after rescaling them to the same starting point, overlap.
That is, if the velocity $v_s$ does not depend on the threshold $\lambda_s$,
then the plot of $S\,'(g) \doteq S(\lambda_s,g) / S(\lambda_s,0)$ should be
universal.
The inset of Fig.~\ref{hKS} reveals however a weak dependence on $\lambda_s$: 
the slopes corresponding to the different thresholds are 0.23, 0.25 and 0.28, 
indicating that $v_s$
ranges in the interval $[0.11-0.14]$. Whether the fluctuations are due to finite
size corrections or are an indication of a whole spectrum of velocities, we
cannot say. The increase of the velocity when $\lambda_s$ decreases suggests,
however, that the {\sl less unstable} directions are characterized by a more
efficient propagation.

Bernoulli maps certainly represent the simplest model for testing the spreading
properties of perturbations in chaotic systems, but the pecularities of the
model (above all, the absence of fluctuations for the local multipliers) may
invalidate the generality of the conclusions. Though this is not true for the scaling
properties of the standard Lyapunov spectrum, it is nevertheless wise to
repeat the analysis above for different models. 

For this reason we now discuss the
case of asymmetric tent maps, for which 
\begin{equation}
f(y) = \begin{cases} 
                     {\displaystyle{
                                      \frac{y}{b}
                                                   }}  & {\rm if}
                                      \quad 0\le y\le b \cr
					\ & \ \cr
                     {\displaystyle{
                                      \frac{1-y}{1-b}
                                                      }}  & {\rm if}
                                      \quad b\le y\le 1. 
\end{cases}
\end{equation}

\begin{figure}[tpb]
\includegraphics*[width=13cm]{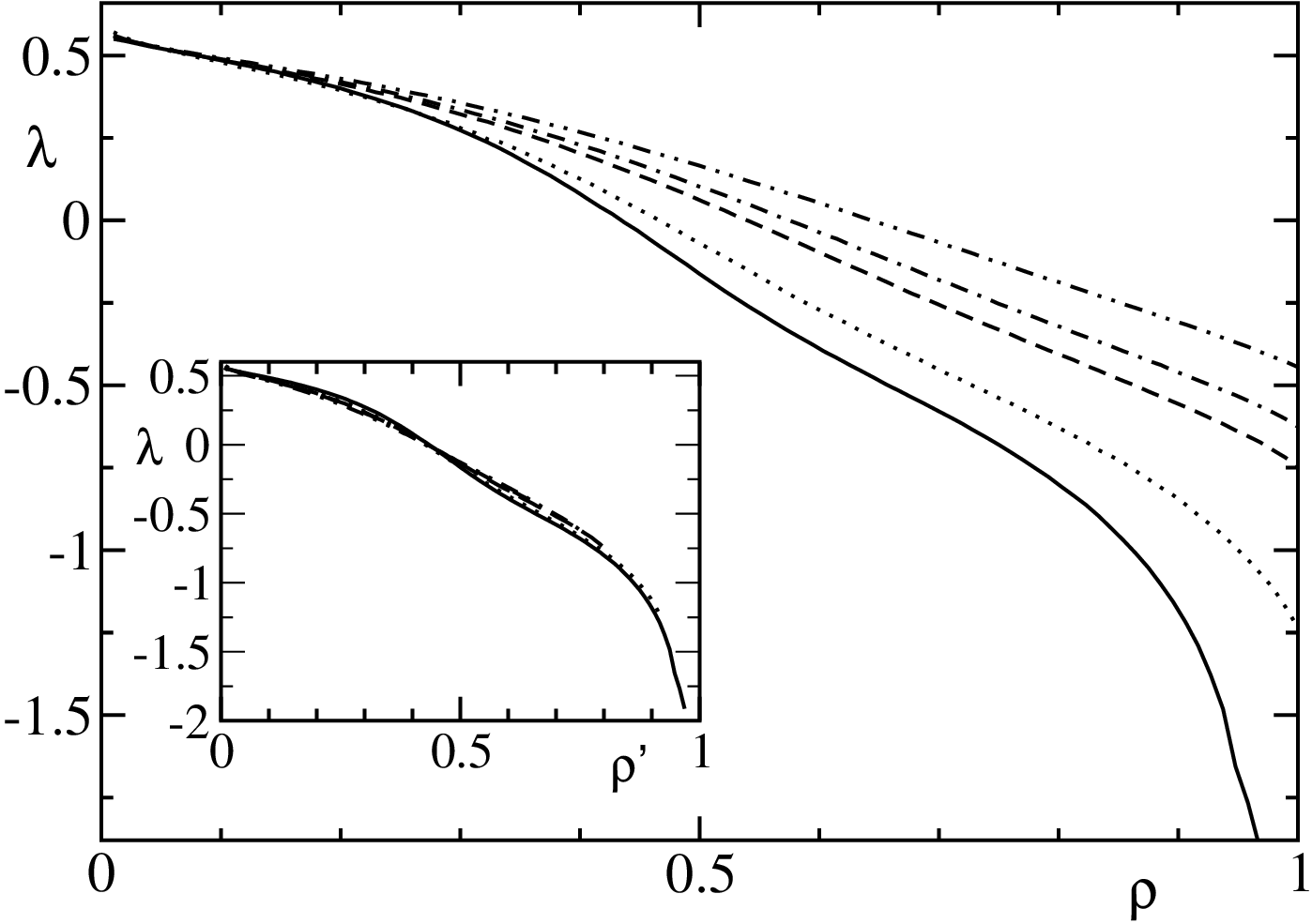}
\caption{Open-system Lyapunov spectra of asymmetric tent maps, for a window
of size $L=96$, with coupling $\eps = 1/3$ and $b=3/4$.
From bottom to top, the curves correspond 
to $g=0$, 1/4, 3/4, 1, and 3/2, respectively. In the inset, the same spectra 
are plotted after suitably rescaling the $\rho$ axis.}
\label{allt}
\end{figure}

\begin{figure}[tpb]
\includegraphics*[width=13cm]{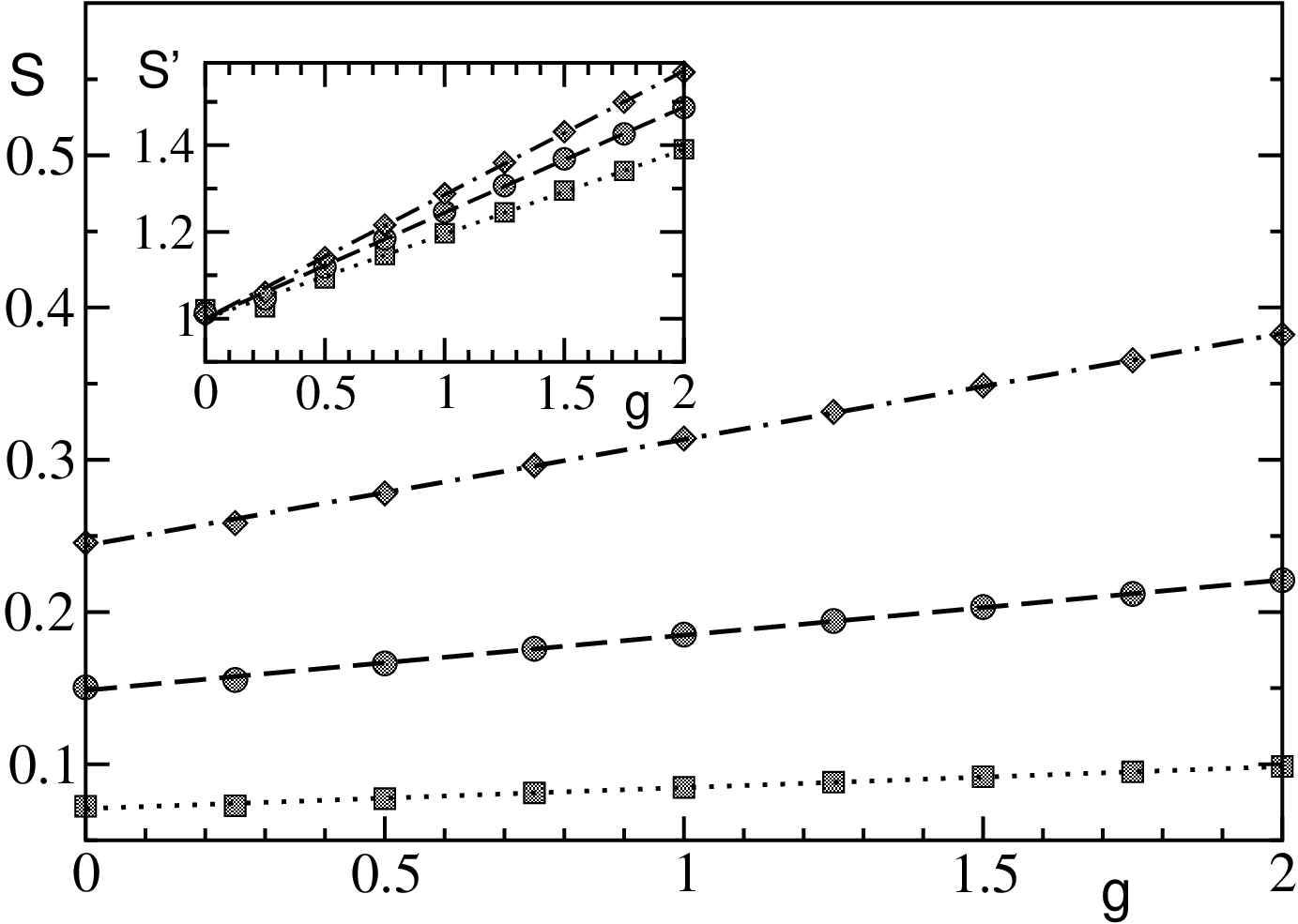}
\caption{The pseudo entropy $S(\lambda_s,g)$, defined in Eq.~(\ref{snodeep}), for the same tent
maps of the previous figure, versus $g$. From top to bottom, it is
$\lambda_s = -0.2,\ 0.0\ {\rm and}\ +0.2$. In the inset
the ratio $S\,' = S(\lambda_s,g) / S(\lambda_s,0)$ is plotted, with the same symbols.}
\label{sogt}
\end{figure}

While the simulations on this model have been performed for several choices of
the coupling $\eps$, and map parameter $b$, the
results here presented refer mainly to the $(\eps , b)  = (1/3 , 3/4)$ case.
Once more, we observe that the OSLS converges to an asymptotic shape when
simulations are performed for increasing time (and thus window size) at a fixed
value of $g$. Some
OSLS's are plotted in Fig.~\ref{allt}, where we again see the same flattening
tendency for the spectrum and a similar collapse, as observed for the Bernoulli maps. 
In order to test quantitatively the (visually suggested) 
hypothetical hyperscaling, we again investigate the behaviour of the pseudo
entropy $S(\lambda_s,g)$ for various choices of the threshold $\lambda_s$. The results
plotted in Fig.~\ref{sogt} confirm what found before, though indicating a more
evident increase in the slopes (in the rescaled representation) of the fits
as $\lambda_s$ is lowered, thus reinforcing the idea that different velocities
are into play\footnote{The simulations performed with a weaker coupling,
$\eps=1/6$, show a qualitatively similar behaviour, though, as expected, with a
slower propagation, $v_{S}=0.05\pm 0.01$.}.  
As before, decreasing the threshold
$\lambda_s$, from $+0.2$ to $0.0$ and to $-0.2$, we observe an increase in the
velocity, $v_{S} = 0.10,\ 0.12,\ 0.14$, respectively, in agreement with the
interpretation, already suggested in the Bernoulli case, of a more efficient propagation
of disturbances along {\it more stable} directions.

Finally, as a prototype of a model with more than one variable per lattice site
and also as an example of a system with conservation of volumes, we have
studied coupled symplectic maps,
\[
p^i (t+1) = p^i (t) - K\left[ \sin{q^i (t)} + \eps \left( T^{i+1}(t) - T^i(t)\right)\right]
\]
\[
q^i(t+1) = q^i(t) + p^i(t+1)\ ,
\]
where $ T^i(t)\doteq \sin{[q^i(t)-q^{i-1}(t)]}$. Consequently, for the evolution 
of perturbations, we have,
\begin{eqnarray*}
\delta p^i(t+1) &=& \delta p^i(t) - K\left[ \cos{q^i(t)} - \eps(D^{i+1}(t) +
D^i(t))\right] \delta q^i(t) + \\
&-& \eps K \left[ D^{i+1}(t) \delta q^{i+1}(t) + D^i(t) 
\delta q^{i-1}(t) \right]
\end{eqnarray*}
and
\[
\delta q^i(t+1) = \delta q^i(t) + \delta p^i(t+1)\ ,
\]
where $ D^i(t)\doteq \cos{[q^i(t)-q^{i-1}(t)]}$.

Again, different values of the parameters have been chosen, though we present 
here just the results related to the case $K=5$ and $\eps=1/6$.

\begin{figure}[tpb]
\includegraphics*[width=13cm]{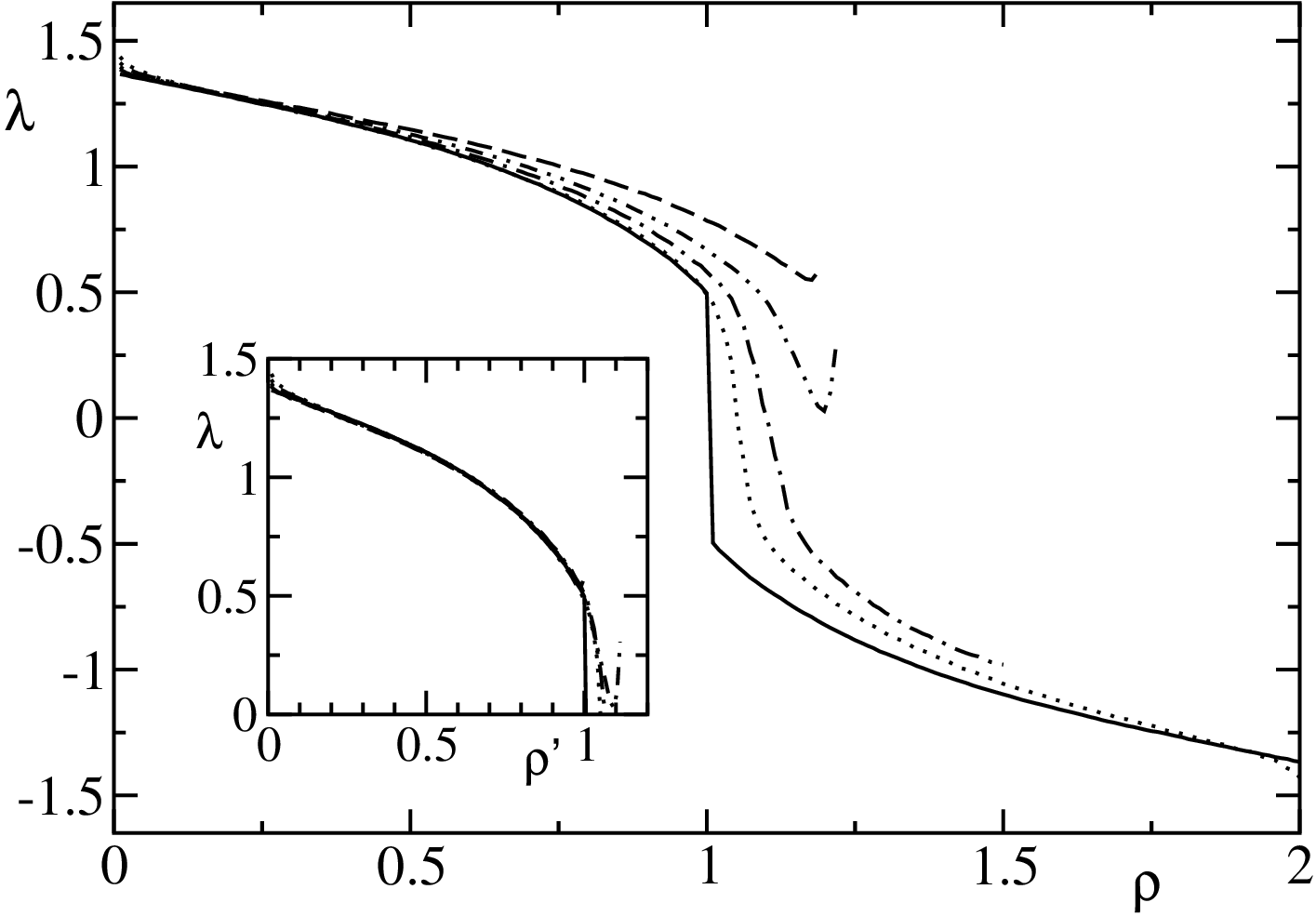}
\caption{Open-system Lyapunov spectra of coupled symplectic maps for $L=96$
({\it i.e.}, $N=192$ exponents).
From bottom to top, the curves correspond to $g=0$, 1/4, 3/4, 1, and 3/2, 
respectively. In the inset, the same spectra 
are plotted after suitably rescaling the $\rho$ axis.}
\label{alls}
\end{figure}

In Fig.~\ref{alls}, the OSLS in the case of a window of size $L=96$ (thus, for
$2L=192$ Lyapunov exponents) are shown, for several values of $g$. The inset in
the same figure points out the good scaling behaviour of these spectra.
Performing on them the same analysis as before, through the computation of
$S\,'$, we find a disturbance propagation velocity $v_s=0.050\pm 0.005$
considerably smaller than in the previous cases, though it should be stressed
that the coupling strength is now smaller by a factor 2.

The velocities obtained through the analysis performed in this section are
basically the volume spreading velocities. As mentioned in the
beginning of this section, another velocity, $v_c$, can be determined from the
zero of the spectrum of the convective Lyapunov exponents. In order to compare
the two velocities, we have computed also $v_c$, using the
procedure described in Ref.~\cite{PTepl94}, to which we refer for more details.
Here we limit to report the results of the application of such a method to the
models discussed in this paper. In Table I, the two sets of velocities
can be mutually compared. For a meaningful comparison with the velocity
obtained from the corresponding convective spectra, we have reported there the
value of $2v_s$, since the effective length implicitly
determined when estimating the Lyapunov spectrum over a time $T$, is the
average length over such an interval and thus it is half of the final length
(if the growth is uniform, as we are assuming). Therefore each velocity has to
be doubled, if we want to make a direct comparison with $v_c$.

\begin{center}
\begin{table}[hbt]
{
\begin{tabular}{||l||c|c||} 
       \hline 
       Model & $2v_s$ & $v_c$ \\ 
       \hline\hline
       Bernoulli $(\eps=1/3)$ &\ $ 0.24\pm 0.03 $\ &\ $ 0.48\pm 0.01 $\ \\ 
       \hline 
       Tent $(\eps=1/3)$ &\ $ 0.24\pm 0.04 $\ &\ $ 0.90\pm 0.01 $\ \\ 
       \hline 
       Tent $(\eps=1/6)$ &\ $\ 0.09\pm 0.02 $\ &\ $ 0.64\pm 0.01 $\ \\ 
       \hline 
       Symplectic $(\eps=1/6)\ \ $ &\ $ 0.09 \pm 0.01 $\ &\ $ 0.83\pm 0.01 $\ \\ 
       \hline 
\end{tabular} 
}     
\caption{Synoptic table containing the velocities determined from the
open-system Lyapunov spectra and from the zero of the convective Lyapunov
spectra for the classes of maps discussed in the text. \label{table}}
\end{table}
\end{center}

We see that in all the cases, the volume spreading velocity is significantly
different from (and smaller than) the propagation velocity of perturbations,
thus indicating that at least two mechanisms exist in spatio-temporal chaos
that contribute to the propagation of information. Indeed, a moment's
reflection reveals a crucial difference between the two velocities. On the one
hand, we see that if the local multipliers change all by the same constant
factor, $r$, the resulting convective Lyapunov spectrum is shifted by $\ln r$
and, as  a consequence, the velocity $v_c$, corresponding to the crossing point
with the $v$-axis, changes. On the other hand, the velocity $v_s$ remains
unchanged. In fact, we believe it is not by chance that $v_s$ turns out to be
approximately the same for the same values of the coupling strength (see the
table): it measures how the spatial coupling forces an ensemble of perturbations
to cover all neighbouring directions. Thus, in principle, it can be defined even 
for a stable system. In a sentence, we could summarize stating that $v_c$ reflects
the (chaotic) features of the local dynamics, whereas $v_s$ exploits the efficiency
of the of the coupling in favouring the propagation of information.

\section{Scaling behaviour of the invariant measure and concluding remarks}

In this last section we exploit the knowledge of the scaling behaviour of the
OSLS to shed light on the corrections to the Kaplan-Yorke formula arising from
the spatial coupling. We start neglecting the integration over the external
degrees of freedom (see Eq.~(\ref{int})). In this approximation, we are
entitled to use equations (\ref{timres}) and (\ref{KYgeneral}), with the
warning that now the Lyapunov exponents do depend on time. Under the assumption
that hyperscaling holds, Eq.~(\ref{KYdens}) still applies, with the length $L$
replaced by its effective value over a time $T$, 
\begin{equation}
\tilde{D}_{KY}(T,L) = d(1 + 2 vg)L =  dL + 2d v T
\label{KYopen}
\end{equation}
In order to independently check the linear dependence of $\tilde{D}_{KY}$ on
$g$, we have studied the behaviour of $d_{KY}=\tilde{D}_{KY}(T,L)/L$, by
integrating the spectra obtained for different values of $g$, in the case of
coupled Bernoulli maps. From the above argument, we expect that
$d_{KY}=d(1+2v_{KY}g)$, where $v_{KY}$ is, in principle, still another velocity
of propagation, though much of the same nature as $v_s$.

\begin{figure}[tpb]
\includegraphics*[width=13cm]{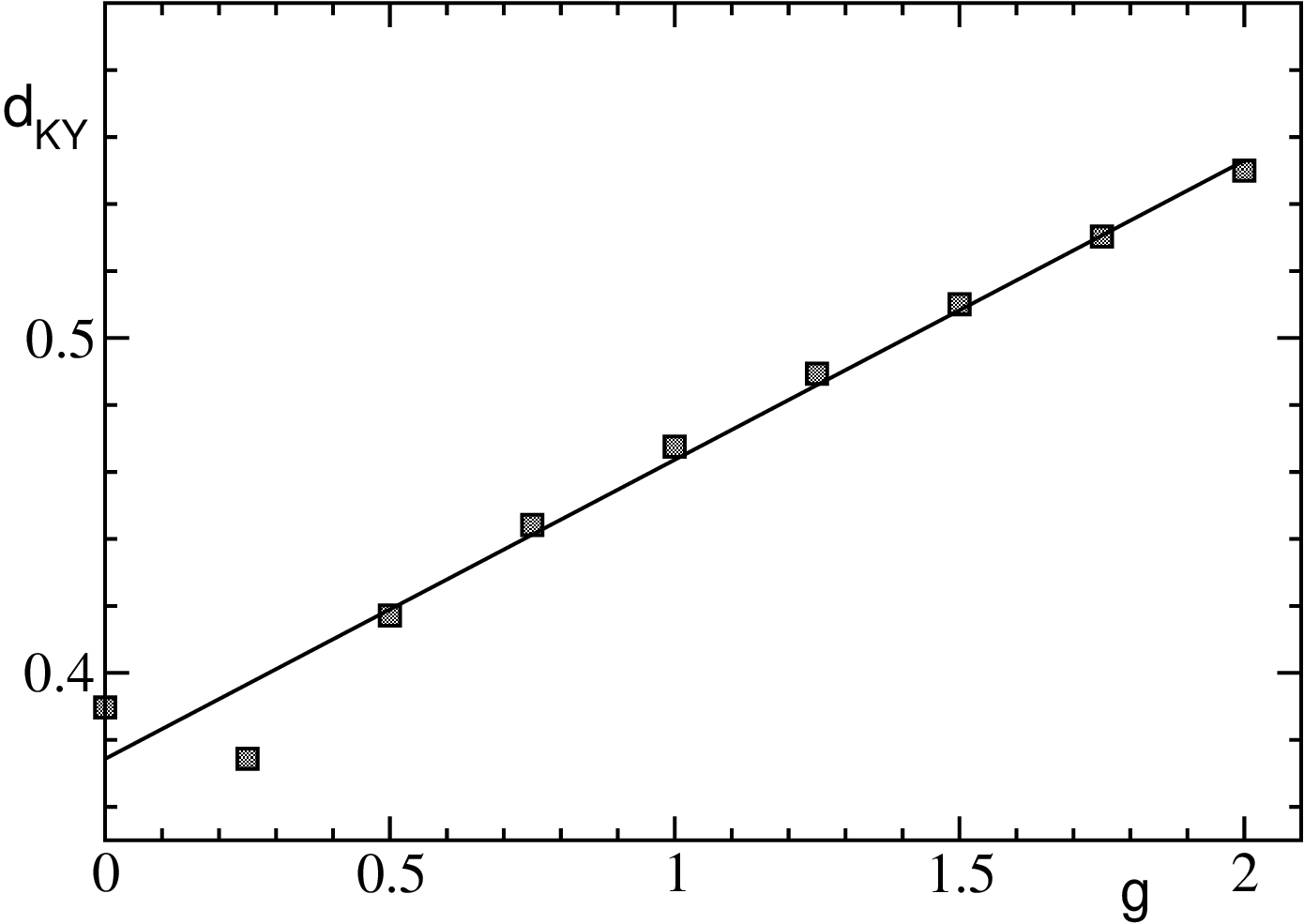}
\caption{The scaled dimension $d_{KY}$ versus $g$, 
as computed from the Kaplan-Yorke formula, for the Bernoulli coupled map lattice.} 
\label{KY}
\end{figure}

The data plotted in Fig.~\ref{KY} indicate indeed a quite clean linear growth.
The deviation of the value corresponding to $g=1/4$ is certainly due to
finite-size corrections, since it has been obtained for the shortest time
compared to the other points, while the datum corresponding to $g=0$
($d_{KY}(0) = 0.39$) is obtained from the standard Lyapunov spectrum (see
Ref.~\cite{IPRT90}), and is therefore not affected by numerical convergence
problems. From the figure, the slope of the curve $d_{KY}(g)$ is 0.09 and from
this (and from $d_{KY}(0)$) it follows that $v_{KY} = 0.12$,  well consistent
with the value, $v_s$, obtained from the generalized KS-entropy approach.

We can now use Eq.~(\ref{timres}) to transform the dependence of $\tilde{D}_{KY}$ 
on time $T$ into a dependence of $D_{KY}$ on the resolution $\eps$. 
Indeed, inverting Eq.~(\ref{timres}), we have
\begin{equation}
T = \frac{\ln \eps}{\lambda_d} \quad  ,
\label{teps}
\end{equation}
where $\lambda_d \doteq \lambda (\rho=d)$.

Inserting then Eq.~(\ref{teps}) in Eq.~(\ref{KYopen}), we obtain
\begin{equation} D_{KY}(\eps,L) = d L + \frac{2 v d}{\lambda_d} \ln \eps
\end{equation} which again reveals a logarithmic dependence on the resolution.
At variance with the past derivations of similar formulas, however, the
multiplicative factor in front of the logarithmic correction follows now from a
well documented discussion of the dynamical evolution. 

However, we should not forget that the above formula represents a lower bound
on the dimension, $D_{KY}$, since we have neglected the integral over ${\bf
x}_\perp^0$ in Eq.~(\ref{int}) that accounts for the uncertainty on the inner
variables induced by the initial lack of knowledge on the outer degrees of
freedom. An accurate quantification of the corresponding propagation of
information is far from trivial, since it may be controlled by nonlinear
mechanisms as it is known to happen in some situations. A typical case, where
this is certainly true is that of the so-called stable chaos~\cite{CLP98}, i.e.
of the irregular behaviour emerging even in the presence of a negative Lyapunov
spectrum. In that case, no extensive contribution to $D$ exists since $d=0$ and
the previously discussed logarithmic correction is absent too. Nevertheless,
finite-amplitude perturbations may enter the window of interest carrying
relevant information.

In order to complement the lower bound above, we present here a heuristic
argument that allows us determining also an upper bound, through an estimate of
the effect of the  disturbances propagating from the boundaries into the region
of observation.  It should be clearly stated that the validity of the following
argument  relies on the assumption that nonlinear mechanisms are negligible. 

From the convective Lyapunov exponents, we know that a perturbation originating
at the boundary at time 0, is amplified by a factor $\exp{\Lambda_c(\ell
/T)T}$, after a time $T$ at a distance $\ell $. Let us assume  that, according
to Eq.~(\ref{teps}), the time $T$ corresponds to a resolution $\ln \eps /
\lambda_d$. Then, the maximum distance travelled by the external perturbation
while still being larger than $\eps$ can be implicitly determined by the
equation
\begin{equation}
  \lambda_d = \Lambda_c(\ell / T) = \Lambda_c \left( \frac{\ell \lambda_d}
{\ln \eps }  \right )\ .
\label{convstand}  
\end{equation}
Let us now $v'$ denote the velocity such that $\Lambda_c(v') = \lambda_d$
(if the whole convective spectrum is larger than $\lambda_d$, $v'$ must be
assumed equal to the maximal possible velocity, which, in the case of nearest
neighbour coupling, is just equal to 1). Eq.~(\ref{convstand}) implies that
\begin{equation}
  \ell = \frac{v'}{\lambda_d} \ln \eps \quad .
\label{ell}  
\end{equation}
By assuming that the distribution of the state variable is independent in
all the $\ell$ sites, one finds that $f\ell $ (where $f$ denotes the number of
variables per lattice site), represents an upper bound to the correction
due to the external degrees of freedom. Accordingly, this correction has again
the same structure as the previous one; altogether, one finds that
\begin{equation}
D_c(\eps,L) \le dL + \frac{2(d v+ fv')}{\lambda_d} \ln \eps .
\end{equation}
Thus, we see that the dependence of the effective dimension on $\eps$ remains
weaker than the $1/\eps$ behaviour deduced indirectly from the treatment in
Ref.~\cite{CE99}. We interpret this as the indication that the rigorous bound
determined in Ref.~\cite{CE99} can be improved. Neverthless, a more detailed
understanding of the propagation of perturbations is also required, in order to
shed further light on the exact structure of the logarithmic corrections. We
are currently exploring the possibility to directly quantify the amplitude of
volume-perturbations in a simplified class of systems where most of the 
calculations can be carried on analytically \cite{CP03}.

The above analysis has been perforemd under the implicit assumption that $|\ln
\eps| < \alpha L$ (where $\alpha$ is a suitable dimensional factor). This
limitation is equivalent to the one encountered in the paper by Collet \&
Eckmann \cite{CE99}, where it is stated that the bound (\ref{GLCE}) applies
only in the region where $\eps > a/L$. The reasons for these limitations are
due to nonlinear effects. In the previous section we have seen that in the
linear approximation any perturbation, initially restricted to $W_L$, sooner or
later diverges along all the $L$ existing directions (inside the window).
However, nonlinear corrections may become important much before the amplitude
of the perturbation becomes of order 1 inside $W_L$. Indeed, we have seen that
perturbations grow outside $W_L$, too; in particular, those that initially
decay inside $W_L$, grow outside the window. As soon as their amplitude
becomes  ${\mathcal O}(1)$ in the outer part, the effect of nonlinearities
cannot be any longer neglected even inside $W_L$, though their amplitude is
therein very small. Since we have seen that high resolutions correspond to
relatively long time, we cannot expect our approach to hold for those tiny
scales where $|\ln\eps|/L$ is no longer small.

It is now worth commenting about the relationship between the open-system
Lyapunov exponents introduced in this paper and the previously devised
chronotopic Lyapunov approach \cite{LPT96,LPT97}. There, it was conjectured
that all linear stability properties of 1d spatially extended dynamical system
can be obtained from the {\it entropy potential}: a function of the spatial and
temporal growth rate of generic perturbations. Although we have not been able
to find the specific link with the OSLS, there is no reason to think that the
information contained in that class of Lyapunov spectra is not contained in 
the entropy potential. Finding the relationship between the two approaches
would be very interesting not only from a conceptual point of view, but also
because it would allow for a much easier computation of the OSLS. We must,
indeed, recall that an accurate determination of Lyapunov spectra in open
systems is hindered by the high accuracy that it requires: this limitation is
certainly crucial in the context of continuous space-time systems such as,
e.g., the complex Ginzburg-Landau and the Kuramoto-Sivahsinsky equations.

Finally, we expect the open system approach to be of some relevance also in
connection with the nonequilibrium dynamics of Hamiltonian systems. For
instance, in Ref.~\cite{G02}, Gallavotti was interested in the expansion rate of
local (in space) volumes. This is nothing but the function $S$, defined by
Eq.~(\ref{snodeep}), for $\lambda_s$ corresponding to the minimum Lyapunov
exponent. 

\section*{Acknowledgements:} One of us (AP) thanks Alessandro Torcini for
illuminating discussions in the early stages of this work.

\end{document}